\documentclass{ws-procs9x6}

\begin{document}

\title{X-ray Luminosity Functions of Active Galactic Nuclei}

\author{Takamitsu Miyaji}

\address{Physics Department, Carnegie Mellon University\\ 
5000 Forbes Ave., Pittsburgh, PA 15213, USA\\ 
E-mail: miyaji@cmu.edu}


\maketitle

\abstracts{
 In this proceedings paper, I overview the current status
of the X-ray luminosity function of AGNs in the soft (0.5-2 keV) band, 
extended using {\it XMM-Newton} and {\it Chandra} survey data.
We found that the number density of low luminosity AGNs peaks
later in the history of the universe ($z\sim 1$) than the 
that of high luminosity AGNs ($z\sim 1.7-3$).  
I also describe the basic results of a spectroscopic followup 
project of a complete {\it HEAO-1} hard X-ray limited sample 
of AGNs using {\it ASCA} and {\it XMM-Newton} and present
separate intrinsic hard X-ray luminosity functions for
unabsorbed and absorbed AGNs. We found that the absorbed AGN XLF 
drops more rapidly at high luminosities, indicating a deficiency
of absorbed luminous AGNs. 
}

\section{Introduction}

 X-ray surveys are practically the most efficient means of finding 
active galactic nuclei (AGNs) over a wide range of luminosity and 
redshift. In order to construct an X-ray luminosity function
(XLF) of AGNs, enormous efforts have been made to follow up 
X-ray sources with optical telescopes to establish their nature 
as AGNs and to measure their redshifts. Now that we have fairly  
complete samples of X-ray selected AGNs over 6 orders of magnitude in
flux, from surveys ranging from all the high galactic latitude sky to 
the deepest pencil-beam fields. These enable us to construct and probe 
luminosity functions over cosmological timescales.

 In this proceedings article, I overview the current progress of 
a few projects related to AGN XLF. Firstly, I overview the results 
of the soft X-ray (0.5-2 keV) luminosity function (SXLF), which is the 
continuation of our previous work with {\it ROSAT} samples\cite{mhs00}, 
extended with deep {\it XMM-Newton} and {\it Chandra} surveys.

 While the soft X-ray surveys select against obscured AGNs, in the current
situation, the number of available objects and area-flux 
coverage from extensive surveys make them useful for probing 
detailed behaviors of XLFs of the unabsorbed portion of the AGN 
activity. A complementary XLF in the hard band (2-10 keV) (HXLF) 
enables us to also look into obscured AGNs, and thus it provides most direct 
measure of the accretion onto supermassive
blackholes (SMBHs). Ueda et al. in this volume covers  
our extensive recent work on HXLF (see also Ueda at al. 2003\cite{ueda1}
for a full description).   

 In this article, we also present the results of our {\it XMM-Newton} 
and {\it ASCA} spectroscopic followup of a complete hard X-ray flux-limited
sample of bright AGNs selected from {\it HEAO-1} catalogs\cite{shino}, 
the basic results of which have been integrated in the Ueda et al's  
HXLF.  
  
\section{AGN Soft X-ray Luminosity Function and Evolution}\label{subsec:soft}
\subsection{The Combined Sample}
\begin{figure}[ht]
\centerline{
\epsfysize=2.2in\epsfbox{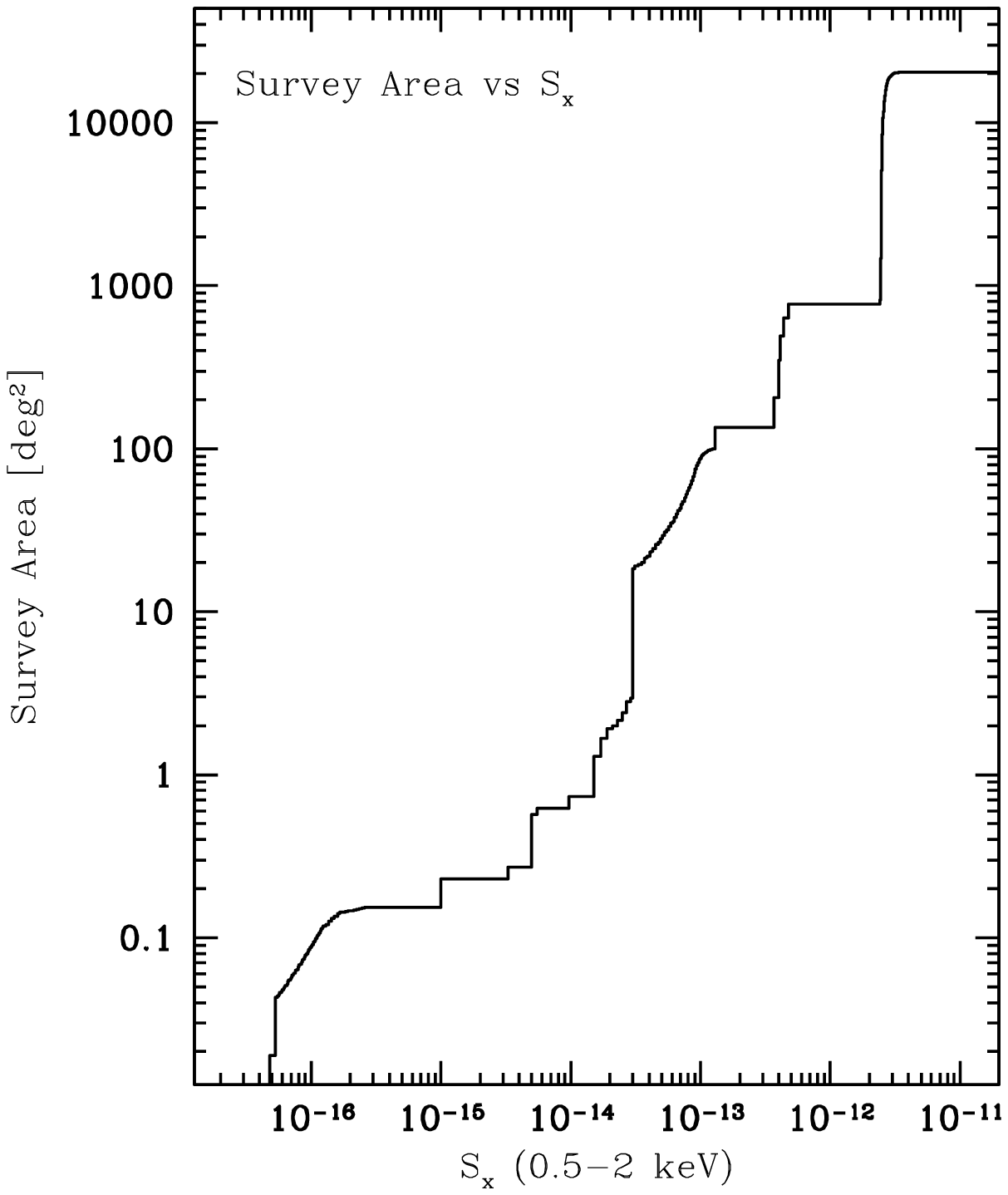}
\epsfysize=2.2in\epsfbox{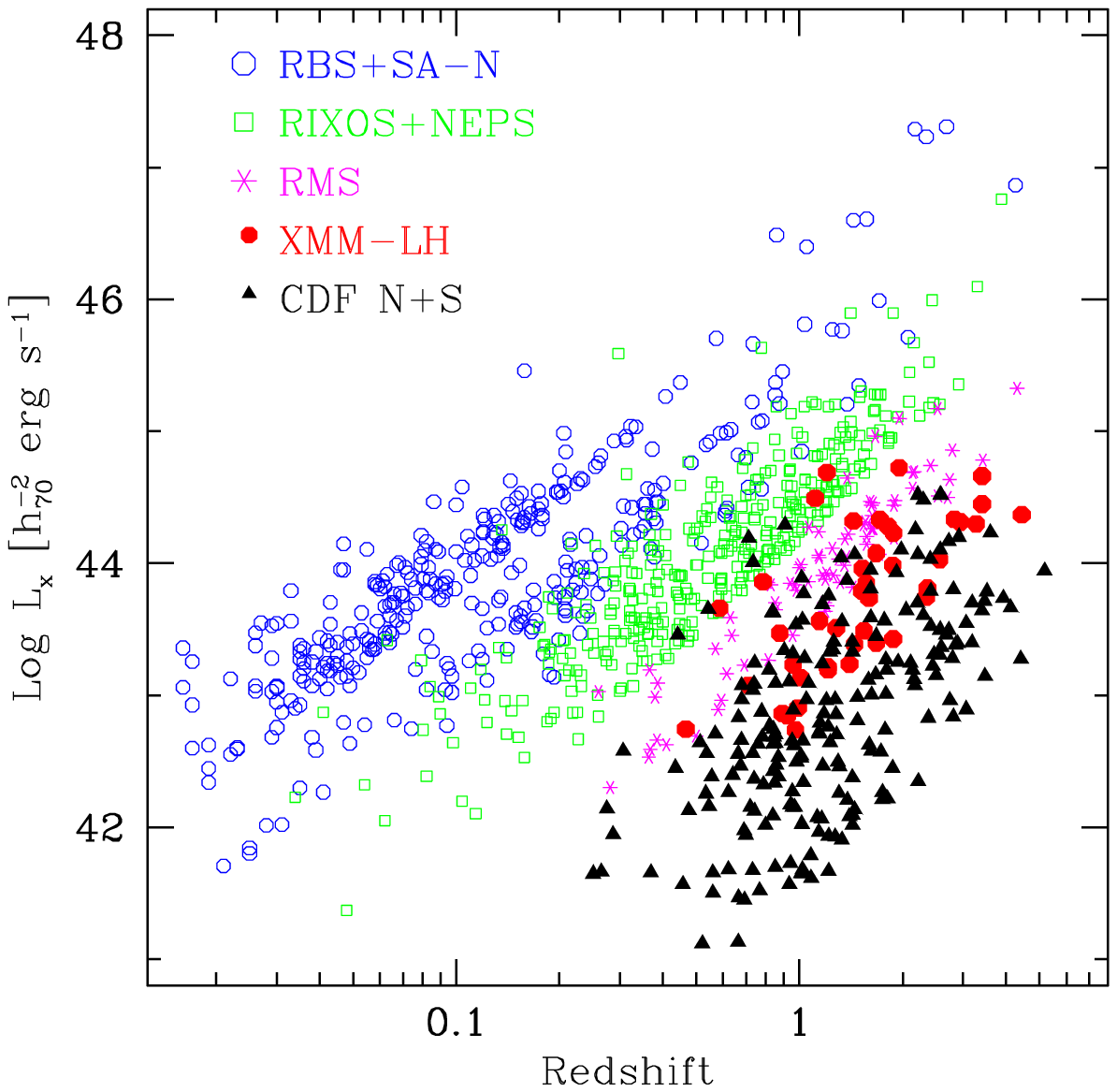}}   
\caption{{\it Left:} The survey area of the combined soft X-ray sample
as a function of flux. {\it Right:} The soft X-ray sample in the 
$z$-${\rm Log}\;L_{\rm x}$ plane. \label{fig:samp}}
\end{figure}

 In addition to the {\it ROSAT} samples used in our previous 
work\cite{mhs00,mhs01}, we have added AGNs from the {\it ROSAT} North
Ecliptic Pole Survey (NEPS)\cite{neps}, from an {\it XMM-Newton} observation
on the Lockman Hole \cite{mainieri} and the {\it Chandra} Deep Surveys 
South\cite{gyula}/North\cite{barger}. For the Lockman Hole region, 
the inner part based on the {\it ROSAT} HRI has been replaced by 
a new XMM-Newton sample\cite{mainieri}. Four medium-deep {\it ROSAT} 
surveys used in our previous work\cite{mhs00,mhs01}, i.e., 
the UK Deep Survey\cite{ukd}, Marano Field\cite{marano}, North Ecliptic
Pole (deep PSPC-pointing)\cite{nep} and the
outer part of the Lockman Hole (PSPC)\cite{rds} 
are now collectively called
the {\it ROSAT} Medium-Sensitivity Survey (RMS). The combined
area versus limiting flux relation and the redshift-luminosity diagram
are shown in Fig. \ref{fig:samp}. The samples are summarized
in Table \ref{tab:samp}. We have tried to limit
our analysis to ``type 1'' AGNs (including narrow-line
Seyfert 1 galaxies). In the {\it ROSAT} samples, we selected 
type 1 AGNs mainly from the optical classifications. For the {\it XMM-Newton} 
Lockman Hole and {\it Chandra} DEEP field samples, the optical classification 
is supplemented by hardness ratios of the X-ray sources. AGNs with 
$z<0.015$ have been excluded from the analysis to avoid the possible 
effects of the local large scale structure. Details will be explained
in Hasinger et al. (in preparation). 
   
\begin{table}[h]
\tbl{The Soft X-ray Sample}
{\footnotesize
\begin{tabular}{@{}cccccccc@{}}
\hline
Survey & Area & $S_{\rm x14,lim}$ & $N_{\rm AGN}$ & Surey & Area & $S_{\rm x14,lim}$ & $N_{\rm AGN}$\\
{} & [deg$^2$] & [cgs]  &   &{} & [deg$^2$] & [cgs]  &   \\[1ex]
\hline
RBS\cite{rbs}& $2\cdot 10^4$ & $\approx 250$ & 203 &
   RMS\cite{ukd,marano,rds}    &  1.-0.5  & .74-.32  &  83\\[1ex]
SA-N\cite{app}  &  684.-35. & 47.-13. & 134 &    
   LH-XMM\cite{mainieri} & .33-.10  & .13-.08  &  42 \\[1ex]
NEPS\cite{neps} &  80.-0.35 & 80.-1.5 & 162 &  
   CDF-S\cite{gyula}  & .09-.02  & .06-.02  & 115 \\[1ex]
RIXOS\cite{rix}  & 19.-16. &  8.5-3.0 & 196 &   
   CDF-N\cite{barger} & .09-.02  &.06-.005  &  97 \\[1ex]
\hline
\end{tabular}\label{tab:samp}
}
\end{table}

\subsection{SXLF and Evolution with Redshift}

\begin{figure}[ht]
\centerline{
\epsfysize=2.2in\epsfbox{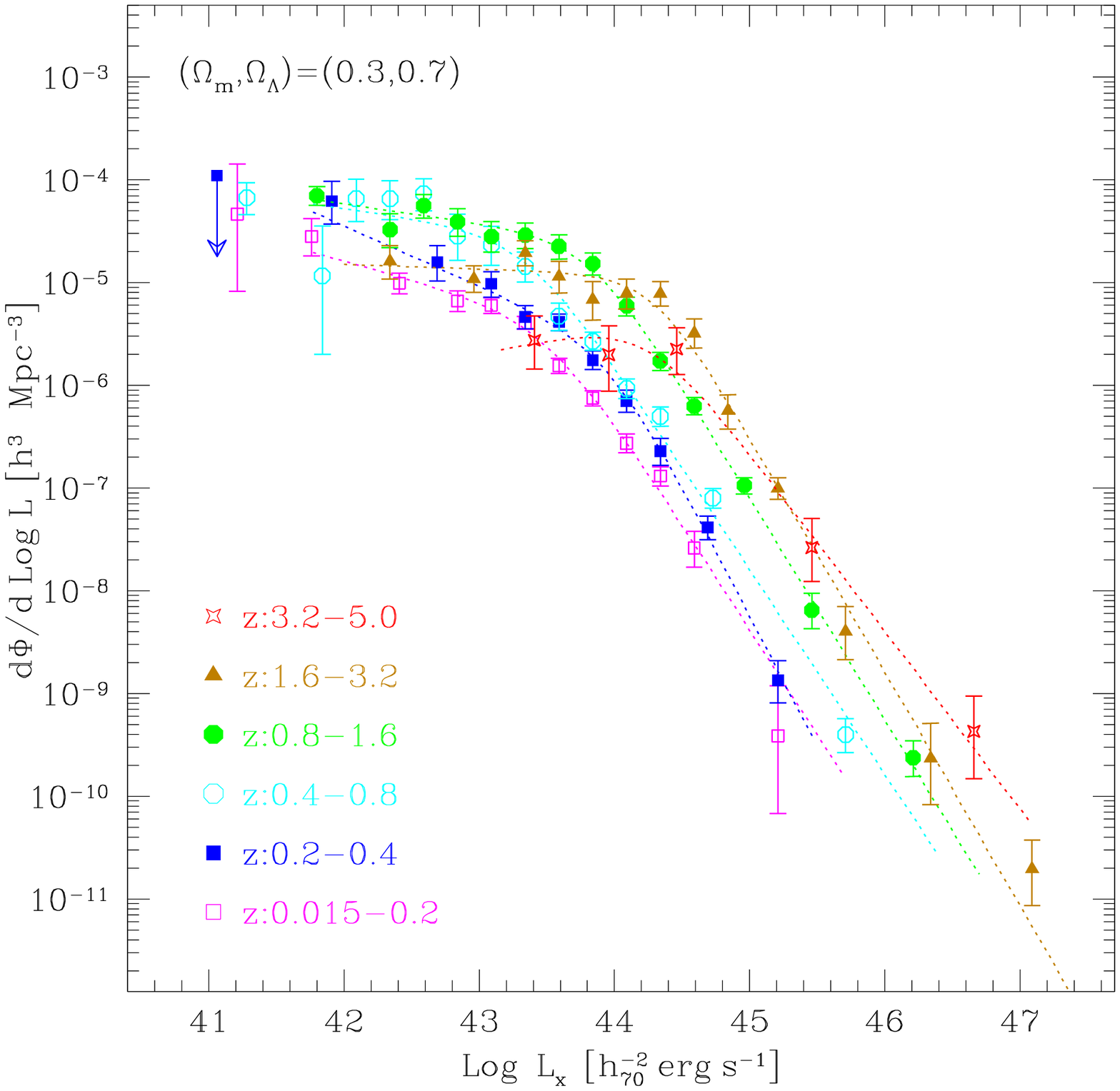}
\epsfysize=2.2in\epsfbox{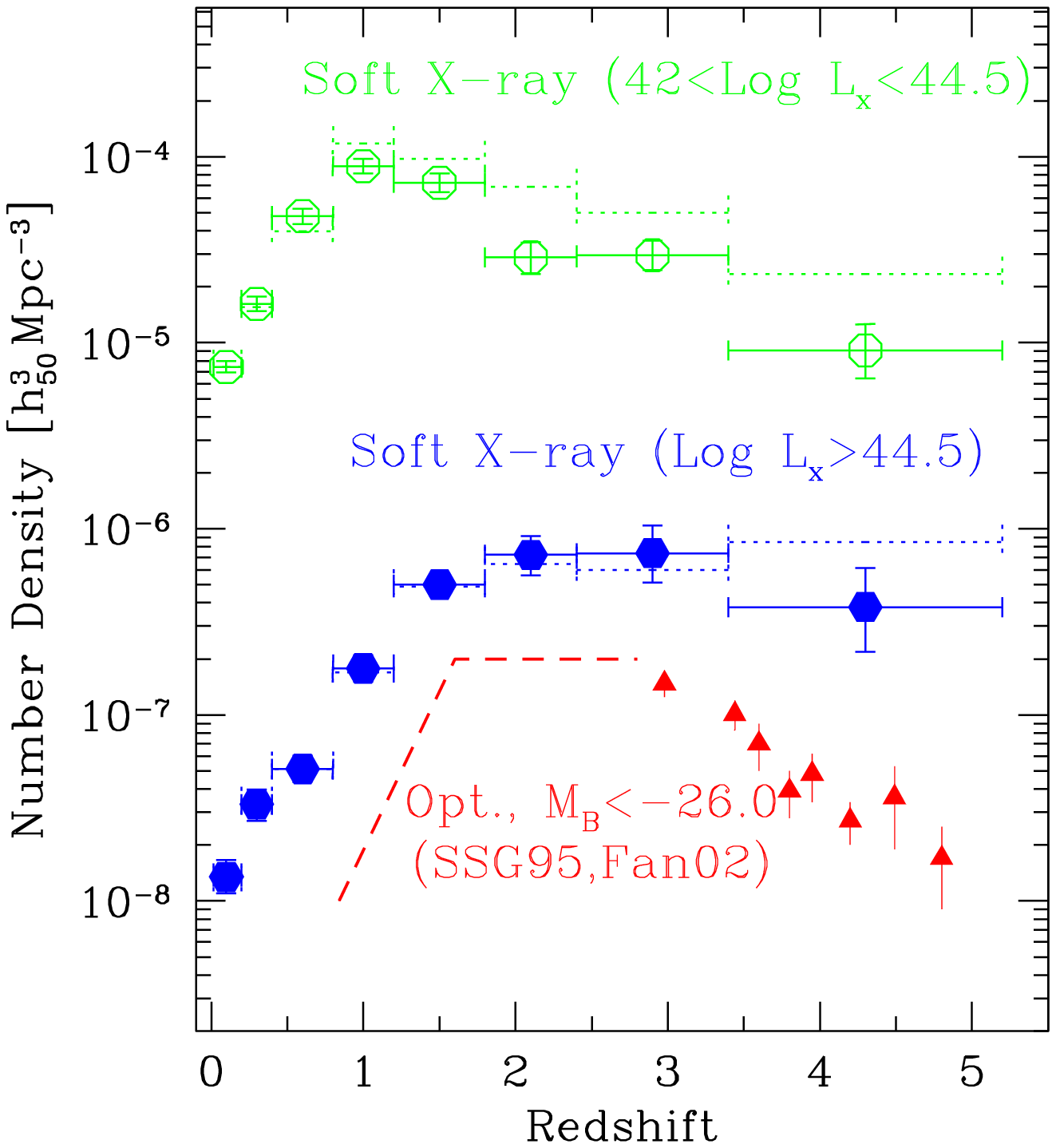}
}   
\caption{
{\it Left:} The SXLF for different redshift bins, calculated
 using the $N^{\rm obs}/N^{\rm mdl}$ method. Errors bars corresponds 
 to Poisson 1$\sigma$ errors.  
{\it Right:} The number densities of soft X-ray selected 
 AGNs with luminosities below and above ${\rm Log}\;L_{\rm x}=44.5$
 ${\rm erg\,s^{-1}}$.
 The same curve for optically-selected QSOs with $M_{\rm B}<-26.0$
 are shown. The $(\Omega_{\rm m},\Omega_\Lambda)=(1,0)$
 cosmology is used in this plot for historical comparisons.}
\label{fig:xlf}
\end{figure}

 The soft X-ray Luminosity functions (SXLF) in different redshift bins
have been calculated using the $N^{\rm obs}/N^{\rm mdl}$ estimator,
where each redshift bin has been maximum-likelihood fitted with a 
smoothed two power-law form and the model value at the center of each bin 
is multiplied by the ratio of the actual number of AGNs in the bin to the 
model-predicted number\cite{mhs01}. Nominally, corrections for 
incompleteness due to unidentified X-ray sources have been made by 
using an {\em effective} survey area, derived by multiplying the 
geometrical survey area by the completeness of the survey (i.e.
identified fraction of the detected X-ray sources). This method is 
valid when sources remain unidentified 
because of random reasons that are not correlated with the 
intrinsic properties of the source (e.g. optical magnitude). 
This is not necessarily true, especially  in the deepest surveys. 
Thus we also calculated the XLF or number density upper bounds, 
where all the unidentified {\it XMM-Newton} and {\it Chandra}
sources are assigned (in duplicate) the central redshift of each bin. 
Figure \ref{fig:xlf} (left) shows the SXLF in different redshift bins 
(plotted only the nominal incompleteness correction case).

 Figure \ref{fig:xlf} (right) shows the AGN number densities as a function
of redshift separately for low and high luminosity AGNs. The 
incompleteness upper bounds (see above) are shown in dotted lines. 
Even in the most extreme cases of the incompleteness correction, 
we see that the number density of the low-luminosity AGNs peaks much
later in the history of the universe ($z\sim 1$) than the
high-luminosity case. This is in the opposite sense to the 
prediction from a analytical model based on the hierarchal merging 
and self-regulated accretion by Wyithe \& Loeb \cite{wyithe}.
According to their prediction, the number density of more luminous AGNs 
peak at later in the history of the universe. On the other hand, a prediction
from a numerical simulation by Di Matteo et al.\cite{dimatteo}, 
where gas density, star formation, and AGN formations are 
assumed to be related in a certain simple way, is consistent 
with the observed trend. 
 
 For comparison, we overplot the redshift evolution of optically-selected 
luminous ($M_{\rm B}<-26$) QSO number density\cite{ssg95,fan}. 
In the high X-ray luminosity bin, we are not still certain (within the 
uncertainties in the incompleteness correction) whether we have detected 
the decline (with $z$) in the number density at $z>2.7$, where the 
densities of luminous optical \cite{ssg95,fan} and radio\cite{shaver} 
QSOs clearly show a drop. Note, however, that when we take a different 
method on incompleteness correction involving optical magnitude limits 
with our soft X-ray sample, a density 
decline at $z>2.7$ is preferred even for the high-luminosity sample 
(Hasinger et al. in preparation).

\section{The Brightest Hard X-ray Sample}

\begin{figure}[ht]
\centerline{
\epsfysize=2.2in\epsfbox{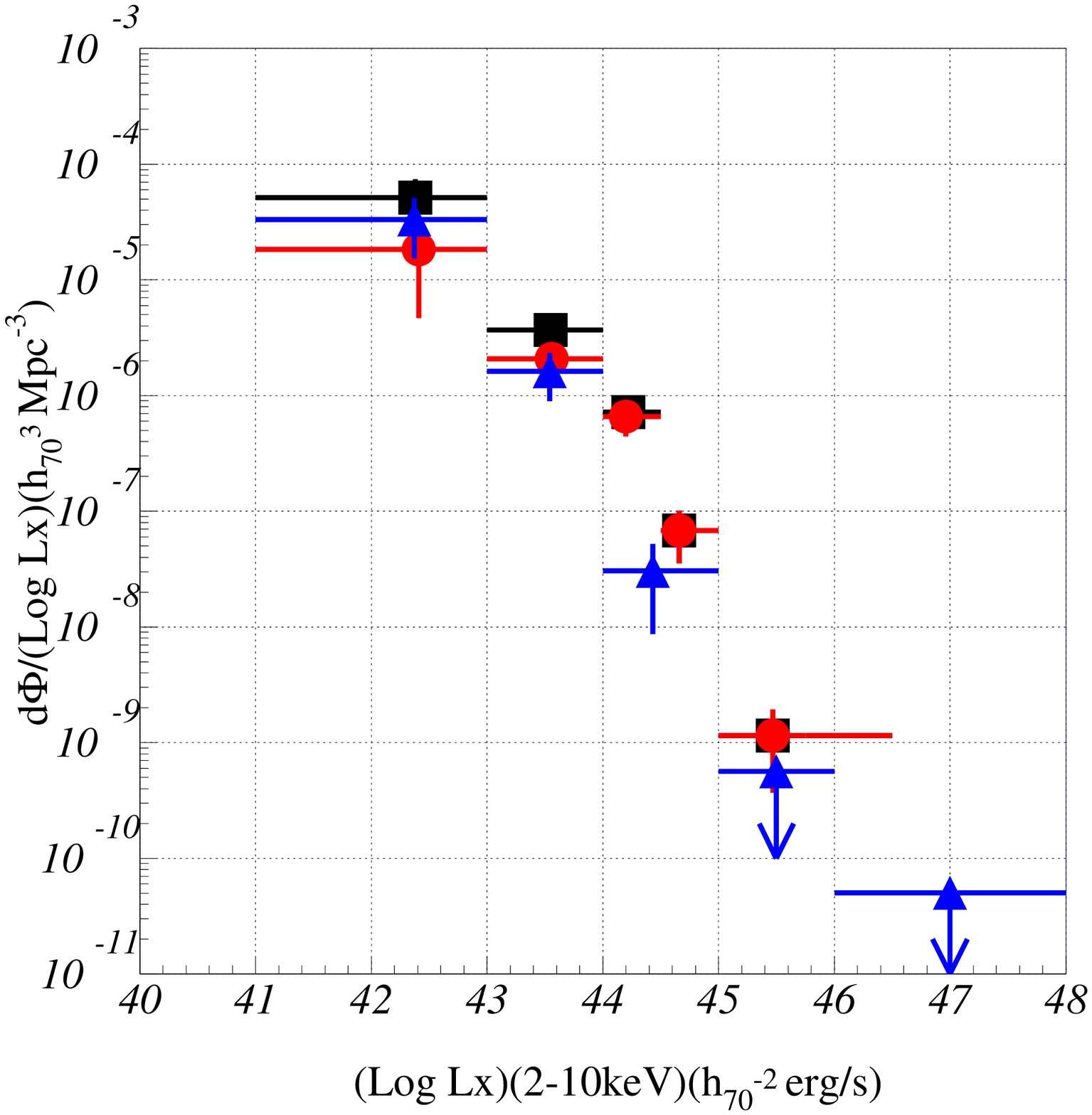}   
\epsfysize=2.2in\epsfbox{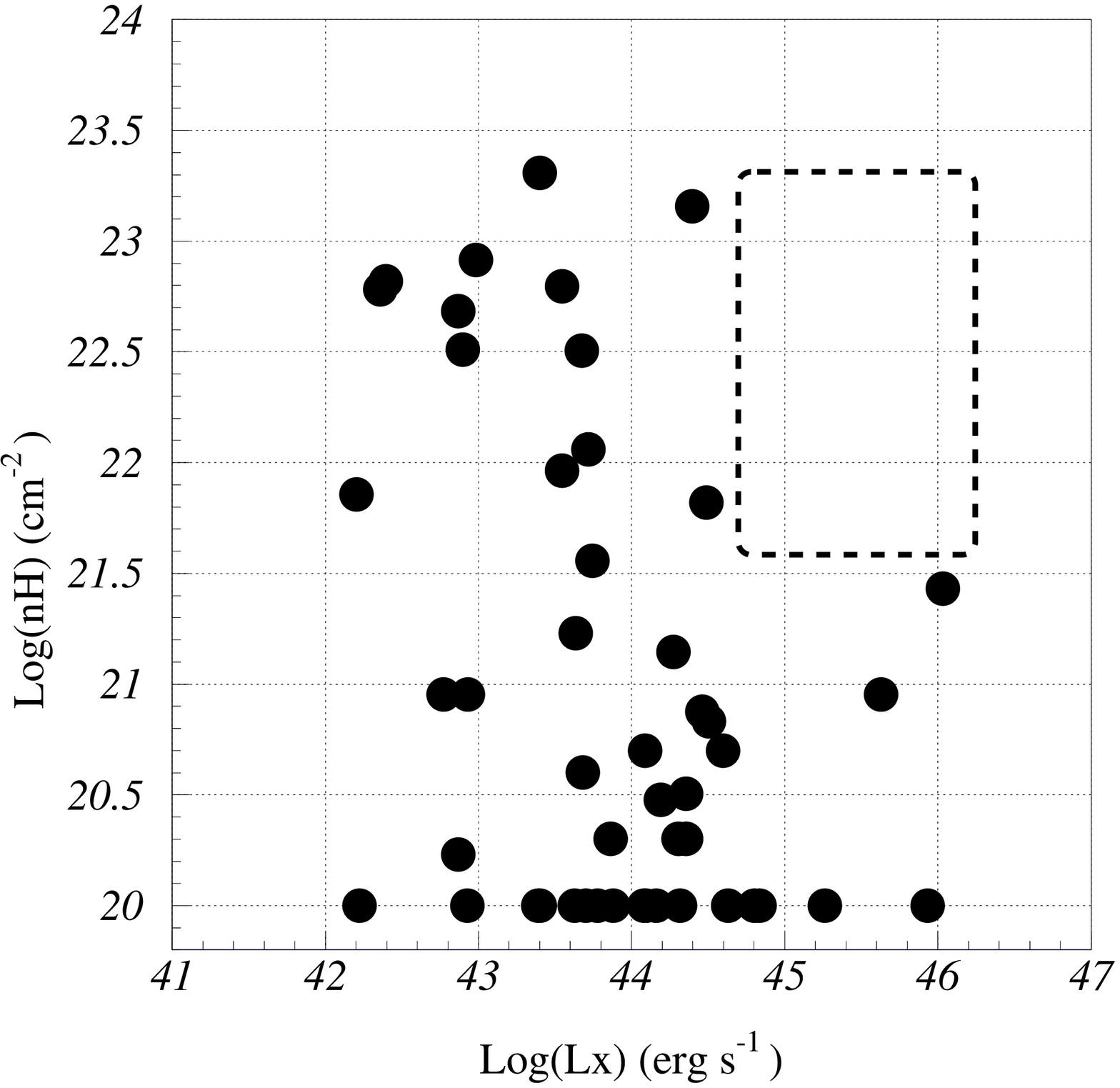}}
\caption{{\it Left:} The local {\em de-absorbed} HXLF (2-10 keV) 
 of unabsorbed (${\rm Log}\; N_{\rm H}\;{\rm cm^{-2}}\leq 21.5$; 
 filled circle) and absorbed ($>21.5$; filled triangle) AGNs. The 
 open squares show the sum of both. {\it Right:} Intrinsic absorption 
 of the sample AGNs are plotted against the intrinsic (de-absorbed) 
 luminosity. There would be $\sim 9$ objects in the region enclosed by 
 a thick dashed line if absorbed and unabsorbed AGNs intrinsic HXLFs 
 were the same besides normalizations.\label{fig:piclma}}
\end{figure}

 In order to construct the luminosity function of AGNs 
in the {\it intrinsic} X-ray luminosity, X-ray surveys 
in the hard band ($E>2$ keV) and X-ray spectroscopic
(or at least hardness) information are crucial. In order to complement
deeper {\it ASCA}, {\it XMM-Newton} and {\it Chandra} surveys
with smaller survey areas, we have defined a bright 
hard X-ray selected sample from the {\it HEAO-1} all-sky surveys. 
In addition to the famous Piccinotti et al.\cite{pic} sample from
the {\it HEAO-1} A2 experiment, we have defined a somewhat deeper 
hard X-ray flux-limited sample of AGNs from the MC-LASS
catalog of X-ray sources from the {\it HEAO-1} A1/A3 experiments
\footnote{http://heasarc.gsfc.nasa.gov/docs/heao1/archive/heao1\_catalog.html} 
in a limited region. The AGN sample from the latter catalog 
was investigated in detail by Grossan\cite{grossan}. As a total,
49 AGNs are defined and we have made spectral analysis of  
all of them (except one) using {\it ASCA} and {\it XMM-Newton}
observation  from archive as well as our own proposals. We have
determined the intrinsic absorption $N_{\rm H}$ and underlying
power-law index $\Gamma$. This enabled us to construct separate local 
HXLFs for absorbed and unabsorbed AGNs as functions of de-absorbed (intrinsic)
luminosity (Fig. \ref{fig:piclma}, left). We see that the absorbed AGN HXLF 
drops more rapidly than the unabsorbed one at high luminosities. This can also
be demonstrated in the $L_{\rm x}$ -- $N_{\rm H}$ plot (Fig. \ref{fig:piclma}, 
right). Suppose absorbed and unabsorbed AGNs had the same HXLF shape, there
would be $\sim 9$ AGNs in the region enclosed by a thick dashed line
in this figure. See Shinozaki et al.\cite{shino}. Full results will be 
reported by Shinozaki et al. (in preparation).

\section*{Acknowledgments}

 I thank my collaborators on the projects described in this article,
especially G\"unther Hasinger, Maarten Schmidt, Keisuke Shinozaki,
Yoshitaka Ishisaki and Yoshihiro Ueda. I thank the conference organizers
for the invitation to give a talk. This work has been supported by the 
NASA LTSA grant NAG5-10875.


\begin{thebibliography}{0}
\bibitem{app} I. Appenzeller et al. {\it A\&A}, {\bf 364}, 443 (2000)
\bibitem{barger} A. Barger et al. {\it AJ}, {\bf 126}, 632 (2003)
\bibitem{nep} R.G. Bower et al. {\it MNRAS}, {\bf 281}, 59 (1996)
\bibitem{dimatteo} T. Di Matteo et al. {\it ApJ}, {\bf 593}, 56 (2003)
\bibitem{fan} X. Fan et al. {\it AJ}, {\bf 121}, 54 (2001)
\bibitem{neps} I.M. Gioia et al. {\it ApJS}, {\bf 149}, 29 (2003)
\bibitem{grossan} B. Grossan PhD Thesis, MIT (1992)
\bibitem{mainieri} V. Mainieri et al. {\it A\&A}, {\bf 393}, 425 (2002)
\bibitem{ukd} I.M. McHardy et al. {\it MNRAS}, {\bf 295}, 641 (1998) 
\bibitem{rix} K.O. Mason et al. {\it MNRAS}, {\bf 311}, 456 (2000)
\bibitem{mhs00} T. Miyaji, G. Hasinger, M. Schmidt  {\it A\&A}, {\bf 353}, 25 (2000)
\bibitem{mhs01} T. Miyaji, G. Hasinger, M. Schmidt  {\it A\&A}, {\bf 369}, 49 (2001)
\bibitem{pic} G. Piccinotti et al. {\it ApJ}, {\bf 253}, 485 (1982)
\bibitem{ssg95} M. Schmidt, D.P. Schneider, J. Gunn {\it AJ}, {\bf 110}, 68 (1995)
\bibitem{rds} M. Schmidt et al. {\it A\&A}, {\bf 329}, 495 (1998)
\bibitem{rbs} A. Schwope et al. {\it AN}, {\bf 321}, 1 (2000) 
\bibitem{gyula} G.P. Szokoly et al. {\it ApJS} submitted (astro-ph/0312324) 
   (2004)   
\bibitem{shaver} P.A. Shaver et al. {\it Nature}, {\bf 384}, 439 (1996) 
\bibitem{shino} K. Shinozaki, T. Miyaji, Y. Ishisaki, Y. Ueda et al. 
  in Proceedings of the "Stellar-Mass, Intermediate-Masss, and Supermassive 
  Black Holes" in press (2004) (astro-ph/0402363)
\bibitem{ueda1} Y. Ueda, M. Akiyama, K. Ohta, T. Miyaji {\it ApJ}, 
  {\bf 598}, 886 (2003)
\bibitem{wyithe} J.S.B. Wyithe , A. Loeb {\it ApJ}, {\bf 595}, 614 (2003)
\bibitem{marano} G. Zamorani et al. {\it A\&A}, {\bf 346}, 731 (1999)
\end{thebibliography}
\end{document}